# Study on Gravitational Waves from Binary Mergers and Constraints on the Hubble Parameter

Rishav Laloo[1], Prachi Gupta[1], Arun Kenath[1*]

[1]Department of Physics and Electronics, Christ University, Bangalore - 560029

[*]email: kenath.arun@christuniversity.in

[*]https://orcid.org/0000-0002-2183-9425

**Abstract:** Einstein's General Theory of Relativity predicted the existence of gravitational waves (GWs), which offer a way to explore cosmic events like binary mergers and could help resolve the Hubble Tension. The Hubble Tension refers to the discrepancy in the measurements of the Hubble Constant, $H_o$, obtained through different methods and missions over various periods. By analyzing gravitational wave data, particularly from mergers that also emit light (electromagnetic radiation), such as Bright Sirens, we aim to reduce this tension. This paper will investigate the properties of GWs produced by these binary mergers and utilize a mathematical framework to tackle the Hubble Tension. Future advancements in gravitational wave astronomy, particularly with initiatives like LIGO-India and LISA, promise to enhance research outcomes. The ground-based LIGO-India will increase sensitivity and improve localization, while the space-based LISA will target lower frequency ranges of GWs, enabling the detection of signals from a wider array of sources.

**Keywords—** Gravitational waves; binary mergers; LIGO; Hubble tension

## 1 Introduction

In 1916, Albert Einstein's General Theory of Relativity introduced the concept that gravity results from the space-time continuum distortion caused by a massive object. This distortion has the potential to propagate throughout the cosmos, leading to the emergence of gravitational waves. Gravitational waves are a result of the non-zero third-time derivative of the mass quadrupole moment of an isolated system's energy-momentum tensor.

If a source is at a large distance, gravitational waves can be expressed as small deviations from flat space-time [1]:

$$h_{\mu\nu} = g_{\mu\nu} - \eta_{\mu\nu} \quad (1)$$

where $\eta_{\mu\nu} = diag(-1,1,1,1)$, $g_{\mu\nu}$ is the metric of the space-time in which gravitational waves propagate and $h_{\mu\nu}$ describes the departure of the full metric from flat space-time. In 1974, observations of the double neutron star system PSR B1913+16 provided the first indirect evidence of gravitational waves. One of the stars in this pair is a pulsar that emits electromagnetic pulses at regular radio frequencies as it rotates. Russell Hulse and Joseph Taylor monitored the pulses from PSR B1913+16 over several years and found that the two stars were orbiting each other at an increasingly rapid pace. The rate at which the stars spiraled closer together aligned with predictions from general relativity with an accuracy of 0.5 percent [2]. This discovery, published in 1978, was the first experimental evidence for the existence of gravitational waves and offered strong support for Einstein's theory of gravity.

On September 14, 2015, LIGO (discussed more in the later sections) detected the first direct evidence of gravitational waves from the merger of two black holes, creating a significant stir in the scientific community. However, this was just the beginning. On August 17, 2017, LIGO detected gravitational waves again, this time resulting from the collision of two neutron stars. This event was particularly notable because it was accompanied by a gamma-ray burst (GRB).

It was recognized that if gravitational waves have an electromagnetic counterpart, such as a gamma-ray burst (GRB), they can serve as standard sirens, allowing for direct measurement of the distance to the source. By combining this distance measurement with the radial velocity obtained from the gravitational waves, it becomes possible to determine the Hubble Constant $H_0$. Thus, gravitational waves have become a valuable tool for determining the Hubble Constant and addressing the Hubble Tension, which refers to the discrepancies in the measured values of the Hubble Constant.

In his General Theory of Relativity, Einstein showed that Newton's law of Gravitation can be expressed in terms of gravitational field. This led to the field equations known today and is given by:

$$R_{\mu\nu} - \frac{1}{2} R g_{\mu\nu} = \frac{8\pi G}{c^4} T_{\mu\nu} \quad (2)$$

(Note: We have excluded the cosmological constant Λ for the time being)

The Einstein Field Equations [3] are ten equations, contained in the tensor equation shown above, which describe gravity as a result of space-time being curved by mass and energy.

- $R_{\mu\nu}$ is the Ricci Tensor and it is a component of the larger Riemann curvature tensor (which is a mathematical notion associated with the non-commutivity of covariant derivatives). The Ricci tensor encodes important information about the distribution of matter and energy in space-time.
- $R$ is the Ricci Scalar which compares the size of any object from Euclidean space to curved space. The Ricci scalar is a single, real number that characterizes the overall curvature of space-time at a specific point.

  In four-dimensional space-time, the Ricci scalar can be defined as:
  $$R = g_{\mu\nu} R^{\mu\nu} \qquad (3)$$
  In practical terms, it quantifies the intrinsic curvature of space-time at a specific point, and it provides a measure of how gravity is affecting that region of space-time
- $g_{\mu\nu}$ is the metric tensor that defines the geometry of space-time in the framework of general relativity. It consists of ten independent components followed by the equation
  $$ds^2 = g_{\mu\nu} dx^\mu dy^\nu$$
  where $ds^2$ describes an infinitesimal length along any curve in any coordinate system and $dx^\mu$ and $dy^\nu$ are displacements in arbitrary coordinates.
- $T_{\mu\nu}$ is the Energy-Momentum tensor which describes the density and flux energy and momentum in space-time and acts as the source of space-time curvature [3].

The exact solutions of the EFE depend on various factors, including the distribution of matter and the geometry of space-time.

One of the key predictions of General Relativity (GR) was the existence of gravitational waves. In GR, mass and energy distort space-time. Massive objects like stars and planets create gravitational fields by bending space-time. The degree of this curvature depends on the distribution of mass and energy in the vicinity. When these objects move and accelerate, the curvature of space-time around them changes. Thus, a sudden alteration in the distribution of mass or energy in a region of space-time generates ripples or waves in the curvature itself. These ripples are known as gravitational waves, and they propagate through space at the speed of light.

# 2  Gravitational Waves

As mentioned earlier, gravitational waves (GWs) are ripples in space-time generated by some of the universe's most violent and intense events. Albert Einstein predicted the existence of gravitational waves in his general theory of relativity in 1916 [4]. Einstein's calculations showed that rapidly moving objects, like neutron stars (NS) or black holes (BH) in orbit around each other, distort space-time, generating waves that propagate outward in all directions. These cosmic waves travel at the speed of light, carrying information about their origins and providing insights into the nature of gravity itself.

In principle, any accelerating physical object generates gravitational waves, including humans, cars, and airplanes. However, the masses and accelerations of objects on Earth are far too small to create gravitational waves strong enough for our detectors to pick up. To detect significant gravitational waves, we must look well beyond our solar system, as the universe contains incredibly massive objects that undergo rapid accelerations, producing detectable waves. Examples include pairs of black holes or neutron stars orbiting one another, as well as a neutron star and a black hole in orbit, and massive stars exploding at the end of their life cycles.

Now, gravitational waves are oscillating disturbances to a flat or Minkowski space-time metric, and are analogous to an oscillating strain in space-time or an oscillating tidal force between free test masses. Like electromagnetic waves, they travel at the speed of light and are transverse, meaning that strain oscillations occur in directions orthogonal to the direction the wave is moving. Unlike electromagnetic waves, gravitational waves are quadrupolar. This means that the strain pattern compresses space in one transverse dimension while expanding it in the orthogonal direction within the transverse plane. Gravitational radiation arises from fluctuating multipole moments in a system's mass distribution. Monopole radiation is forbidden by the conservation of mass, while the conservation of linear and angular momentum rules out gravitational dipole radiation. As a result, quadrupole radiation is the lowest allowed form and is typically the dominant type observed.

In general relativity, gravitational radiation is fully described by two independent, and time-dependent polarizations, $h_+$ and $h_\times$ reflecting the fact that they are rotated 45 degrees relative to one another.

The amplitude of the wave is a different story. The true answer necessitates some difficult computations, however, we can state that the dimensionless strain, $h$ of a wave formed by some mass $M$ moving at some speed $v$ and detected at distance $D$ is roughly [5]

$$h \sim \frac{GM}{c^2} \frac{1}{D} \left(\frac{v}{c}\right)^2 \qquad (4)$$

## 3 Phenomenology of Gravitational Waves

The most intense sources of gravitational waves are those with the highest compactness. Since neutron stars and black holes are the most compact objects in the Universe, they are also the strongest emitters of gravitational waves. The interaction of gravitational radiation with these systems typically influences their dynamics. This interaction either leads to instability, causing exceptionally bright events such as the merging of compact binary stars, or it promotes stability by reducing asymmetrical motion in the system, as seen with rotating neutron stars.

Astronomers have classified gravitational waves into four types based on the object or system that generates them: Continuous, Compact Binary Inspiral, Stochastic, and Burst. Each type of object produces a distinct set of gravitational wave signals that researchers can look for in LIGO data.

### 3.1 LIGO

The Laser Interferometer Gravitational-Wave Observatory (LIGO) is a large-scale physics experiment designed to detect cosmic gravitational waves and advance gravitational-wave astronomy. LIGO consists of three specialized Michelson interferometers located at two sites: the Hanford site in Washington, which has two interferometers—the 4 km-long H1 and the 2 km-long H2—and an observatory in Livingston Parish, Louisiana, which contains the 4 km-long L1 detector. Apart from the shorter length of H2, the three interferometers are fundamentally identical.

While LIGO can detect and analyze gravitational waves independently, its capabilities for astrophysical research can be significantly enhanced by operating within a larger network. Only a network of detectors can accurately determine the direction of travel and capture the complete polarization information of the waves. Over the past decade, a global network of gravitational wave observatories has developed. Notable additions include the Japanese TAMA project, which built a 300 m interferometer near Tokyo; the German-British GEO project, which constructed a 600 m interferometer in Hanover; and the European Gravitational Observatory, which established the 3 km-long Virgo interferometer near Pisa, Italy.

During its first observing run (O1) [6], which took place from September 12, 2015, to January 19, 2016, LIGO made the first-ever direct detection of gravitational waves. The gravitational wave named GW150914 won the discoverers the Nobel Prize.

The second observing run (O2) took place from November 30, 2016, to August 25, 2017, after an update to the detectors. Advanced Virgo joined this run on August 1, 2017. On April 1, 2019, Advanced LIGO and Advanced Virgo commenced their third observing run (O3), which lasted for about a year. The Gravitational Wave Transient Catalog (GWTC-1) includes 11 confident detections (10 binary black hole mergers and 1 binary neutron star merger) along with 14 marginal triggers, all based on the analysis of data from O1 and O2.

This catalog contains notable events such as the first detected event, GW150914, the first three-detector event, GW170814, and the binary neutron star (BNS) coalescence, GW170817 [6]. This was the first time that gravitational and electromagnetic waves had been seen from the same source, providing a unique account of the physical processes at work during and after the merging of two neutron stars.

The third observing run, O3, took place from April 1, 2019, to April 21, 2020. This run focused exclusively on data collection from LIGO and Virgo and was divided into two phases: O3a, which ran from April 1 to October 1, 2019, and O3b, which lasted from November 1 to March 27, 2020 [7].

## 3.2 Binary Mergers

The LIGO and Virgo Collaboration detected the first gravitational waves on September 14, 2015, ushering in a new era in gravitational wave astronomy. September 14, 2015, marked just the beginning. Since then, LIGO has detected several black hole binaries, with more discoveries anticipated. These observations offer insights into the most extreme relativistic conditions and shed light on the existence of black holes with masses ten times that of our Sun. Additionally, by observing two of these massive black holes spiraling around each other at relativistic speeds, we can test whether Einstein's laws of physics continue to hold.

Merging black holes is but one of the expected signals. Other kinds of gravitational waves observed so far come from merging binary neutron stars and neutron star–black hole mergers. In General Relativity, two orbiting objects steadily spiral together as a result of gravitational radiation's loss of energy and angular momentum and eventually merge.

## 3.3 Binary Black Holes

If the two bodies are black holes, they form a single perturbed black hole that radiates gravitational waves as a superposition of quasinormal ringdown modes. Typically, one mode dominates soon after the merger, and an exponentially damped oscillation at a constant frequency can be observed as the black hole settles to its final state, characterized by the combined masses of the original black holes. This combined mass is referred to as the chirp mass and is defined by the following formula:

$$\mathcal{M} = \frac{(m_1 m_2)^{\frac{3}{5}}}{M^{\frac{1}{5}}} \quad (5)$$

where $m_1$ and $m_2$ are the masses of the two black holes and $M = m_1 + m_2$ is the total mass of the system [8]. The chirp mass is the mass parameter that, at the leading order, drives the frequency evolution of gravitational radiation in the inspiral phase.

The rate at which the frequency of the gravitational wave increases as the binary system gets closer to merging is given by [8]

$$\frac{df}{dt} = \frac{96}{5} \pi^{\frac{8}{3}} \left(\frac{G\mathcal{M}}{c^3}\right)^{\frac{5}{3}} f^{\frac{11}{5}} \quad (6)$$

where $f$ is the frequency of the GW, $G$ is the Universal Gravitational constant and $c$ is the speed of light in vacuum.

GW150914 was detected in both LIGO detectors [9] with a 7 ms time difference and a combined matched-filter signal-to-noise ratio (SNR) of 24. At the time of GW150914, only the LIGO detectors were operational, as the Virgo detector was undergoing upgrades, and GEO 600, while active, was not in observational mode and lacked the sensitivity to detect this event. The fundamental characteristics of the GW150914 signal indicated that it was produced by the merger of two black holes. The best-fit template parameters from the search agreed with comprehensive parameter estimation, which identified GW150914 as a near-equal mass black hole binary system with source-frame masses $36^{+5}_{-4} M_\odot$ and $29^{+4}_{-4} M_\odot$ at 90% credible level. The signal grows in frequency and amplitude in around 8 cycles over 0.2 s, from 35 to 150 Hz, where the amplitude hits a maximum.

## 3.4 Binary Neutron Stars

Neutron stars (NS) are astronomical objects with densities roughly of the order of $10^{17} - 10^{18}\ kgm^{-3}$ [10]. Several models have been proposed to constrain the equation of state (EoS) of the neutron star's interior. More than a hundred EoS candidates have been proposed since the prediction [11] and discovery of neutron stars. However, just a few have been realistic and successful in correlating the observations.

A binary neutron star merger was found to be the cause of a gravitational wave signal of about 100 seconds duration that occurred on August 17, 2017, less than 50 Mpc from Earth [12]. This was less than two years after the direct discovery of gravitational radiation from the merging of two 30 $M_\odot$ black holes. The signal was named GW170817. What separated this event from the other merger events was that it marked, for the first time that a cosmic event was viewed in both gravitational waves and electromagnetic waves.

The Fermi and INTEGRAL satellites for high-energy astrophysics separately detected a brief gamma-ray burst in the same sky region, which proved to be related to the gravitational event. The host galaxy was identified as NGC 4993 and Hubble's constant was inferred in a novel manner [13] by combining the cosmological redshift of NGC 4993 and the luminosity distance estimated from GW170817.

Merging binary neutron stars were quickly recognized to be promising sources of detectable gravitational waves, making them a primary target for ground-based interferometric detectors.

Assuming that both NSs may be approximated as point masses, a circular binary orbit decays at a rate given by, $\frac{da}{dt} = -a/\tau_{GW}$ [14].
where $a$ is the binary separation and the gravitational radiation merger timescale $\tau_{GW}$ is given by:

$$\tau_{GW} = \frac{5}{64} \frac{a^4}{q(1+q)M_1^3} = 2.2 \times 10^8 q^{-1}(1+q)^{-1} \left(\frac{a}{R_\odot}\right)^4 \left(\frac{M_1}{1.4M_\odot}\right)^{-3} yr \quad (7)$$

Where $M_1, M_2, M = M_1 + M_2$ are the individual NS masses and the total mass of the binary, respectively, and $q = M_2/M_1$ is the binary mass ratio. The luminosity of such systems in gravitational radiation is [14]

$$L_{GW} = \frac{-dE_{GW}}{dt} = \frac{32}{5} \frac{M_1^2 M_2^2 (M_1+M_2)}{a^5} \quad (8)$$

which, at the end of a binary's lifetime, when the components have approached within a few NS radii of each other, is comparable to the luminosity of all the visible matter in the universe ($\sim 10^{53} erg/s$).

The resulting strain amplitude observed at a distance $D$ from the source (assumed to be oriented face-on) is given approximately by [14]

$$h = \frac{4M_1 M_2}{aD} = 5.53 \times 10^{-23} q \left(\frac{M_1}{1.4 M_\odot}\right)^2 \left(\frac{a}{100 km}\right)^{-1} \left(\frac{D}{100 Mpc}\right)^{-1} \quad (9)$$

at a characteristic frequency

$$f_{GW} = 2 f_{orb} = \frac{1}{\pi} \sqrt{\frac{M}{3}} = 194 \left(\frac{M}{2.8 M_\odot}\right)^{\frac{1}{2}} \left(\frac{a}{100 km}\right)^{-\frac{3}{2}} Hz \quad (10)$$

## 3.5 Black Hole-Neutron Star Binaries

Of the three main types of binaries detectable through ground-based gravitational wave observations, black hole-neutron star (BH-NS) mergers remain the most elusive. Black hole-neutron star binaries are thought to form from two supernovae in a massive binary system. After their formation, the orbital separation gradually decreases due to the long-term effects of gravitational radiation, leading to an adiabatic inspiral motion. Eventually, this process results in the merger of the two objects, forming a black hole system [15]. The lifetime of a binary in a quasi-circular orbit is approximately given by

$$\tau_{GW} = \frac{5c^2}{256 G^3} \frac{r^4}{(M_{BH} M_{NS}) M_{BH} M_{NS}} \approx 1.34 \times 10^{10} yrs \left(\frac{r}{6 \times 10^6}\right)^4 \left(\frac{M_{BH}}{6 M_\odot}\right)^{-1} \left(\frac{M_{NS}}{1.4 M_\odot}\right)^{-1} \left(\frac{M}{7.4 M_\odot}\right) \quad (11)$$

where $r, M_{BH}, M_{NS}$ are the orbital separation, masses of the BH and NS, respectively, and $M = M_{BH} + M_{NS}$.

## 3.6 Why is the merger of BH-NS binaries important?

The merger of BH-NS binaries (more specifically, tidal disruption of a NS by a BH) is physically and astrophysically an important phenomenon, and deserves a detailed study, because of at least three reasons as follows [16].

- The orbital frequency at tidal disruption is highly dependent on the compactness of the neutron star. Therefore, the gravitational waves emitted during tidal disruption can offer vital information about the radius and equation of state (EoS) of the neutron star $(GM_{NS}/R_{NS} c^2)$.

  The masses of the neutron star and black hole will be calculated by analyzing data from gravitational waves released during the inspiral phase. If the neutron star's radius could be calculated or constrained by observing gravitational waves emitted during tidal disruption, the resulting relation between mass and radius of the observed neutron star might be used

- to constrain the EoS of high-density nuclear materials. Therefore, the gravitational-wave observation for BH-NS binaries will provide a new tool for exploring high-density nuclear matter, which is independent of standard nuclear experiments.
- A tidally-disrupted neutron star may form a disk or torus of mass larger than $0.01 M_\odot$ with the density $\geq 10^{11}\ g/cm^3$ and the temperature $\geq 10\ MeV$ around the remnant black hole, if a tidal disruption occurs outside the ISCO. A system consisting of a spinning black hole surrounded by a massive, hot, and dense torus has been proposed as one of the likely sources for the central engine of a GRB. In the merger of BH-NS binaries, the resulting disk is typically compact, with a mass of the order of about $0.1 M_\odot$.
- Material ejected from a tidally disrupted neutron star may be important for understanding the observed abundances of the heavy elements that are formed by rapid neutron capture. One crucial question is whether a fraction of material can escape from the system because the situation is not well prepared for the mass ejection. Tidal disruption of a neutron star occurs typically at an orbital separation of $\sim 10\ GM_{BH}/c^2$.

For a test particle of mass $m$ in a circular orbit around the black hole with $r \sim 10\ GM_{BH}/c^2$, the total energy is approximately $GM_{BH}m/2r = 0.5mc^2$. For a free nucleon, $0.5mc^2 \approx 50\ MeV$. Thus, the issues are, specifically, to answer whether it is possible to give about $50\ MeV$ energy to each nucleon and, if possible, to clarify the relevant process.

## 4 GWs as a possible resolution to Hubble Tension

The standard cosmological model known as the Λ–Cold Dark Matter Model offers a reasonably satisfactory explanation for a significant portion of existing data. Nonetheless, as technology advances and experimental sensitivity increases, deviations from this standard model might become more anticipated [17]. One of these deviations is observed in the value of the Hubble constant $H_0$, which represents the expansion rate of the universe.

The Hubble constant can be determined through two primary methods. The first involves analyzing the cosmic microwave background (CMB) radiation, as done by the Planck mission, by measuring its anisotropies and other properties that provide insights into the early universe. The second approach estimates the Hubble constant by utilizing the distance-redshift relationship for nearby galaxies. This method relies on the luminosities of standard candles, such as Cepheid variables, Type Ia supernovae, the tip of the red giant branch (TRGB), and similar astrophysical indicators.

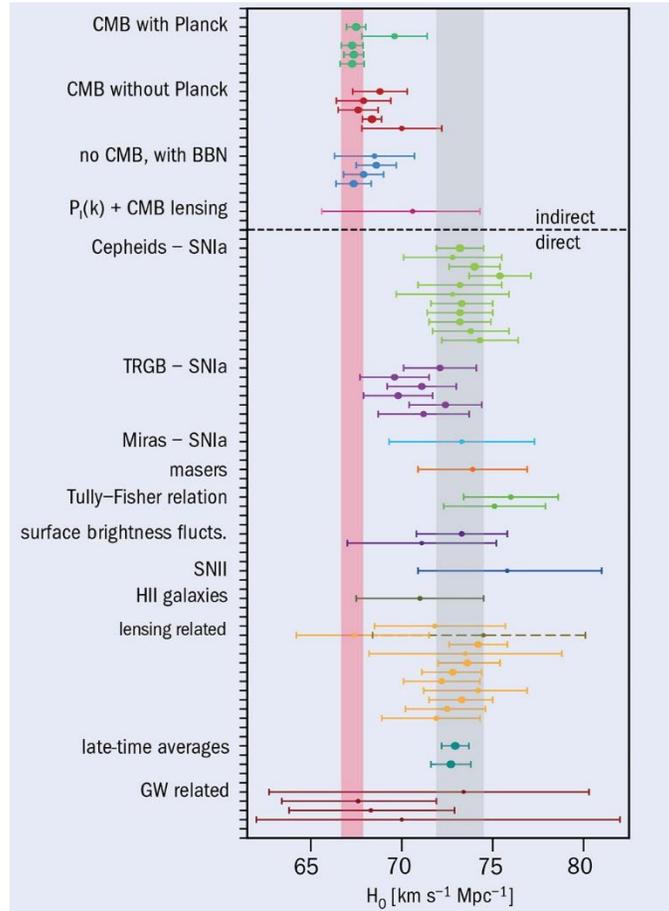

**Figure 1:** *Whisker plot of the Hubble constant through measurements by different astronomical missions and groups* [17].

Figure 1 illustrates the varying estimates of the Hubble constant obtained from different missions over the years. The value inferred from the Planck mission, based on observations of the cosmic microwave background, is approximately $67\ kms^{-1}Mpc^{-1}$, whereas measurements using distance indicators in the local universe yield values in the range of $73 - 74\ kms^{-1}Mpc^{-1}$.

Additionally, an independent determination of the local Hubble constant, derived from the calibration of the Tip of the Red Giant Branch (TRGB) applied to Type Ia supernovae, provides an estimate of $69.8\ kms^{-1}Mpc^{-1}$ [18].

The late Universe measurements give a value of Hubble constant that is significantly different from that measured from the CMBR. It is evident from these observational results that this discrepancy in the measured value of the Hubble constant is not due to observational errors. Hence, we need a theoretical framework to explain this apparent inconsistency.

On 17 August 2017, GW170817 was observed by both LIGO and Virgo, coming from the merger of a binary neutron-star merger. Within 2 seconds of this observation, a Gamma Ray Burst (GRB 170817A) was detected from the region of the sky which was consistent with the LIGO-Virgo observation. This multi-messenger observation enables us to use the signal as a standard siren to measure the Hubble Constant.

Standard sirens do not require any form of cosmic distance ladder and serve as the gravitational wave analog to an astronomical standard candle, enabling the measurement of the Hubble constant. From the Hubble Law [19]:

$$v_H = H_0 d \qquad (12)$$

where $v_H$ is the recessional velocity of the source and is measured using the redshift of the electromagnetic signal and $d$ is the proper distance measured using the GW signal.

Using GW170817 observation, Hubble velocity of the source galaxy NGC 4993 was calculated as $v_H = 3017 \pm 166 \; km/s$ with 68% uncertainty. Distance $d$ was calculated as $43.8 \; Mpc$, resulting in the value of Hubble constant as $69.8 \; kms^{-1}Mpc^{-1}$.

## 5  Modelling

We have used the following relations to derive the Hubble Constant for the various GW events that have been recorded so far:

- **Chirp Mass** is the effective mass of a binary system, in the context of the Quadrupole Gravitational Radiation emitted by it. For two masses $m_1$ and $m_2$, the Chirp Mass is given by

$$\mathcal{M} = \frac{(m_1 m_2)^{\frac{3}{5}}}{(m_1 + m_2)^{\frac{1}{5}}} \qquad (13)$$

- The **Frequency** of gravitational waves at an instant $t$

$$f = \frac{5^{\frac{8}{3}}}{8\pi} \left(\frac{G}{c^3}\right)^{-\frac{3}{8}} (t_c - t)^{-\frac{3}{8}} \mathcal{M}^{-\frac{5}{8}} \qquad (14)$$

where $t_c$ is the time after the merger of the two bodies

- The **Strain** produced due to the passing Gravitational Waves is given by

$$h = \frac{4G\mathcal{M}}{d_L c^2} \left(\frac{Gf\pi\mathcal{M}}{c^3}\right)^{\frac{2}{3}} \qquad (15)$$

where $d_L$ is the luminosity distance to the binary system

- **Hubble Constant**, which governs the expansion rate of the Universe is given by

$$H_o = \frac{cz}{d_L}\left[1 + \frac{z}{2}(1 + q_o)\right] \qquad (16)$$

Here, $z$ represents the redshift and $q_o = -0.55$ is the deceleration parameter. We examined several gravitational wave events detected during the three observational runs (O1-O3) of the LIGO-Virgo collaboration [20]. Our analysis focused on key events, including *GW170817*, the only bright siren event observed thus far, *GW170104*, a binary black hole (BBH) merger, *GW190814*, another BBH event with a high mass ratio, and the double binary black hole system *OJ287* [21]. Using our independent distance relation model, we computed the Hubble parameter ($H_o$) for these events, with the results presented in Table 1.

Additionally, we determined the strain ($h$) and frequency ($f$), for each event and generated plots illustrating their variation with the Chirp Mass, drawing from the LIGO data release encompassing all recorded events thus far. Our results include a plot depicting the Hubble parameter derived from the data release, complete with error bars.

| Table 1: GW events/systems | | | |
|---|---|---|---|
| **GW170817** | **GW170104** | **GW190814** | *OJ287* |
| $H_o = 75 kms^{-1}Mpc^{-1}$ | $H_o = 42.865 kms^{-1}Mpc^{-1}$ | $H_o = 67.74 kms^{-1}Mpc^{-1}$ | $H_o = 74.047 kms^{-1}Mpc^{-1}$ |
| $h = 1.29 \times 10^{-21}$ | $h = 3.47 \times 10^{-23}$ | $h = 2.69 \times 10^{-22}$ | $h = 2.4 \times 10^{-16}$ |
| $f = 297.43$ Hz | $f = 1813.704$ Hz | $f = 651.019$ Hz | $f = 5.316 \times 10^{-9}$ Hz |

From the LIGO data, we made a plot of Frequency vs. Chirp Mass and Strain vs. Chirp Mass for all the events recorded thus far, as depicted in Figure 2. We can see that as the Chirp Mass increases, the frequency decreases as is clear from equation 14.

LIGO's sensitivity lies at roughly 100 Hz to about 10,000 Hz, but the detection of GW signals also depends on the frequency of the incoming GW signal. LIGO has the highest sensitivity in the range of about hundreds of hertz [22].

According to equation 15, larger chirp mass produces higher strain, which can be inferred from Figure 2(b).

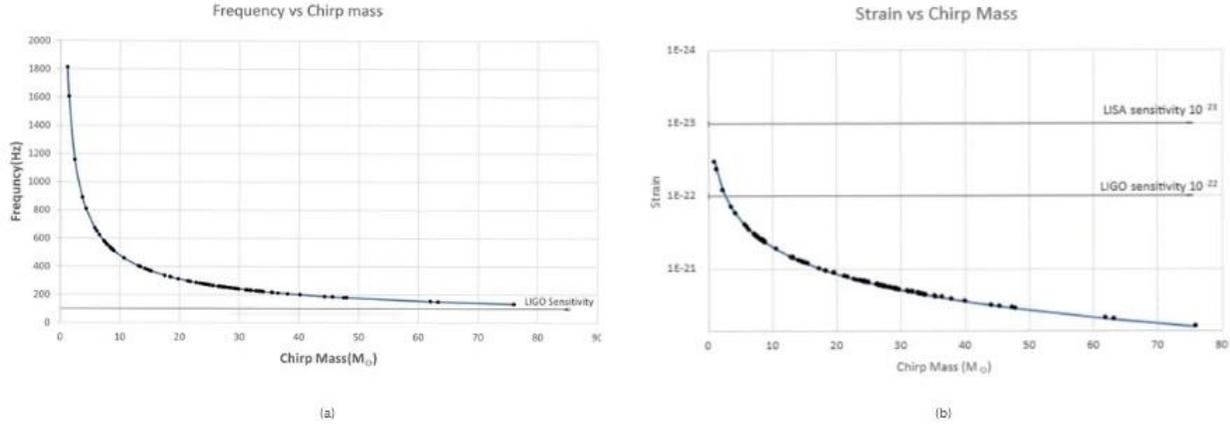

**Figure 2:** *(a) Frequency variation for GW events recorded during the three LIGO observations runs O1, O2, and O3. The lower limit for GW frequency for LIGO is 100Hz as depicted. The upper limit is not shown because LIGO's sensitivity varies with the incoming GW signal. (b) Strain variation for GW events recorded during the three LIGO observations runs O1, O2, and O3. Sensitivity for LIGO and LISA are also depicted.*

According to equation 15, larger chirp mass produces higher strain, which can be inferred from the figure. The sensitivity limit for LIGO and LISA are shown above which, the GW signals cannot be detected. The following Whisker plot (Figure 3) shows the Hubble Constant calculated for all the GW events released by LIGO during its three observation runs.

Hubble Constant with positive and negative error bars using Independent Distance Relation for the detected GWs by LIGO observation runs O1, O2, and O3

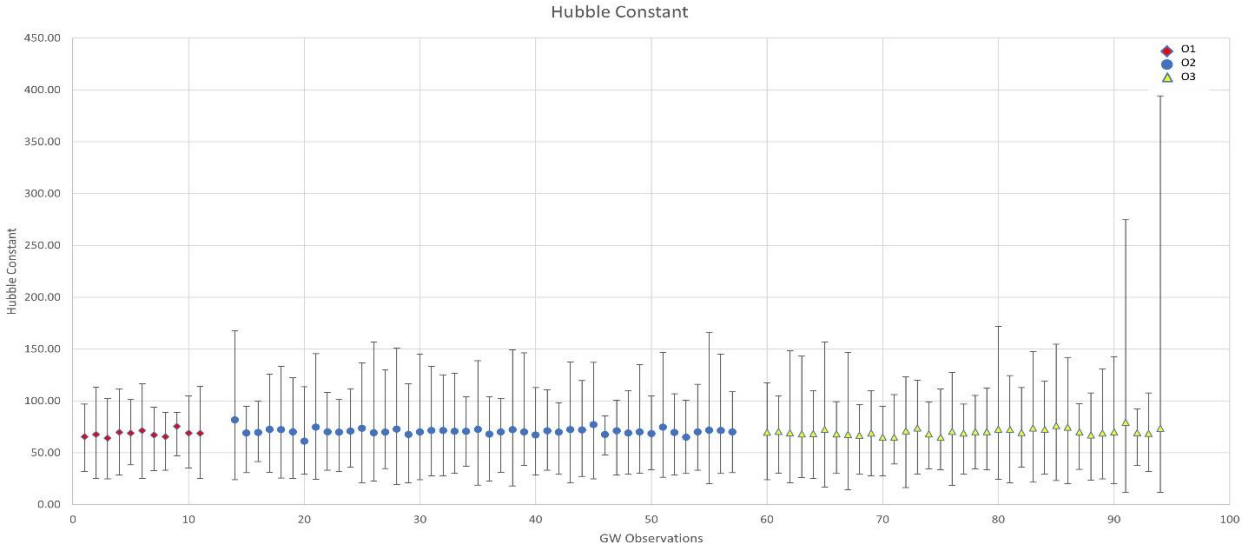

**Figure 3:** *Hubble Constant with positive and negative error bars using Independent Distance Relation for the detected GWs by LIGO observation runs O1, O2, and O3*

In Figure 3, the dots show the central value that was calculated, and the vertical lines show the errors. It is no surprise that we encountered such high errors because all of these events except one (GW170817), are dark sirens and hence, the distance could not be calculated directly and accurately.

## 6 Conclusions

This work has revealed that gravitational waves from binary mergers provide a distinctive way to gather insights about their origins and measure the Hubble constant innovatively, bypassing the traditional distance measurement methods. This technique greatly minimizes measurement errors.

A major challenge is that most detected gravitational wave sources are dark sirens, which do not emit light. This makes it difficult to accurately locate these sources, leading to errors in distance measurements that impact our calculations of the Hubble constant. With the current limited number of interferometers, we cannot effectively resolve this localization issue. Therefore, we need more gravitational wave detectors to enhance our ability to pinpoint these sources.

Furthermore, gravitational wave events with electromagnetic counterparts, the bright sirens, offer greater potential for precise Hubble constant measurements, necessitating detectors sensitive enough to detect these events.

Since the direct detection of gravitational waves in 2015, nearly a century after Einstein's prediction, gravitational wave astronomy has flourished as a promising research area. The LIGO-Virgo collaboration alone has observed over a hundred events stemming from compact binary mergers.

While gravitational waves offer a potent approach for addressing the Hubble Tension, there exist numerous models and hypotheses attempting to explain it. However, currently, there is no method solely relying on gravitational waves to resolve the Hubble Tension without incorporating electromagnetic waves [23].

With only one bright siren detected so far—from the merger of neutron stars in 2017—it is clear that we do not have enough data to effectively tackle the Hubble Tension. To make significant strides in resolving this discrepancy, we need to detect more bright sirens.

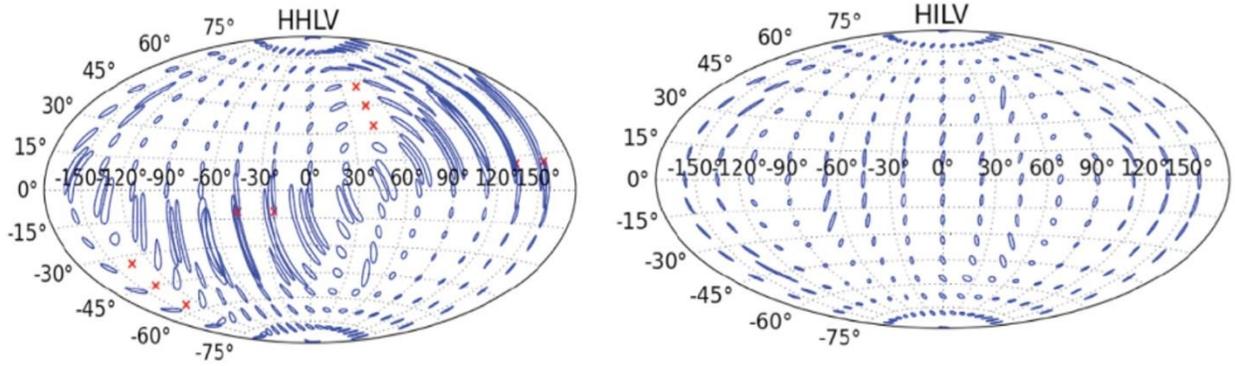

**Figure 4:** *Left: Localization capability: LIGO + Virgo only; Right: Localization capability: LIGO + Virgo+ LIGO-India* [24].

      The current network of gravitational-wave detectors (like LIGO and Virgo) has a limited sensitivity. This means they can only detect strong gravitational waves from relatively nearby events. For more distant mergers, the signal may be too faint to pinpoint the location accurately (see Figure 4). Characteristics inherent to the binary system, such as the existence of supplementary material or an uneven explosion during the merger, can introduce ambiguities in interpreting the gravitational wave signal. Consequently, this may impact the precision of determining the inferred source location. LIGO-India along with the upcoming third-generation ground-based interferometers (such as Cosmic Explorer and Einstein Telescope) and space-based gravitational wave detectors (LISA) will provide more precise cosmological probes by opening up a wider range of GW sources [24].

      Moreover, the issue of the Hubble tension will be thoroughly investigated, allowing us to determine definitively whether the tension arises from inaccuracies in our models or from errors in distance calibrations. Given these considerations, the future will embrace a multi-faceted approach in which astronomers combine data from gravitational wave detectors with observations from a variety of telescopes and observatories, including optical and X-ray instruments. This integration aims to provide a more comprehensive understanding of binary merger events and their environments. This holistic method operates independently of measurements like the CMBR (Planck estimate) or late Universe measurements of the Hubble constant, making it a promising avenue to address the Hubble Tension.

While General Relativity has been highly successful, especially with the direct detection of gravitational waves, it has also revealed certain limitations and issues. These shortcomings prompt theorists to question whether it is the ultimate theory of gravity. Thus, in addition to General Relativity, modified gravity theories (the Finite Range Gravity (FRG) [25], being one of the examples), are also employed to determine the Hubble constant. This is mainly because modified gravity theories often predict additional gravitational wave polarizations compared to standard general relativity, which leads to variations in the response functions of interferometers [26]. So, in turn, the study of Gravitational Waves can provide a way to test the limits of GR and set constraints on any alternate models.